# The solution of the Einstein Equation in the Interior of the Black Holes by using Arbitrary Distribution Functions

N.A. Hussein*, D. A. Eisa** and T. A. S. Ibrahim***

*Mathematics Department, Faculty of Science, Assiut University, Assiut, Egypt.

**Mathematics Department, Faculty of Science, Assiut University, New Valley, Egypt.

***Mathematics Department, Faculty of Science, Elminia University, Elminia, Egypt.

# Abstract

The aim of this paper is to obtain the solution of the Einstein equation in the interior of the black holes by using arbitrary distribution functions; corresponding to Gaussian, Rayleigh, Maxwell-Boltzmann and non-Gaussian distributions. Also we calculate the Hawking temperature, the mass and heat capacity for cosmological horizon and the black hole horizon.

## I. Introduction

A black hole is defined as a place where *gravity* has gotten so strong that the *escape velocity* is faster than light speed. In particular, we have the BTZ black hole (the lower dimensional black holes)[1]. It's helpful to understanding the holographic description of asymptotically anti-de Sitter (AdS) space-times [2]. However the lower-dimensional analog of black hole is not known, but there are solutions with the cosmological horizon [3-4]. If we have available a lower-dimensional analog which could exhibit the key features of its' higher-dimensional counterparts; and the conventional black holes are a perturbative solution in that their horizon sizes scale with the Newton's constant $G$.

For Gaussian after a lot of works there exist a new horizon in the black hole space-time, but anisotropic i.e., $p_r \neq p_\theta$ with energy-momentum tensor $T^\alpha_\beta = diag\left(\rho, -p_r, -p_\theta, -p_\varphi\right)$. For Gaussian not be required always, though beyond the Gaussianity has not been studied much so far. But Myung and Yoon have constructed deformed $AdS_3$ black hole which based on a Rayleigh distribution which is one of the non-Gaussian distributions [5]. Our aim of this paper to report that the four dimension black hole solution when non-Gaussian smearing of point matter are consider exist generally. We solve Einstein equations and discovery $g_{\alpha\beta}$ we find that the high degree of non-linearity possessed by equation so that the solution not be easy, but this problem become easier if we solve it under some assumption [6]. The first exact solution was obtained by K. Schwarzschied in 1916; he thought that the metric static spherically symmetric in empty space time surrounding some spherically massive object like a star. In this paper, we assume that the metric static spherically symmetric but we solve this equation by assuming that the distribution function in general form which including the Gaussian, Rayleigh and Maxwell-Boltzmann distribution ($n = 0,1,2$) respectively. We find that the black holes solution for all higher moment matter distribution except the conventional Gaussian one. This is a new way to constructing black holes solution from hair.

## II. The distribution function of Black Holes

The equations of motion are given by Einstein field equation [7]

$$R_{\alpha\beta} - \tfrac{1}{2}g_{\alpha\beta}R + \Lambda g_{\alpha\beta} = 8\pi G T_{\alpha\beta} \quad (1)$$

Where $R_{\alpha\beta}$ is the Ricci tensor and $g_{\alpha\beta}$ is the metric tensor, $R$ is the scalar tensor, $T_{\alpha\beta}$ is the energy momentum tensor and $\Lambda = \frac{1}{l^2} = -8\pi P$ is the positive cosmological constant and $l$ is the radius curvature. In order to solve equation (1) we assume that the line element take the form

$$d\tau^2 = f(r)dt^2 - \frac{1}{f(r)}dr^2 - r^2 d\theta^2 - r^2 \sin^2\theta\, d\varphi^2 \quad (2)$$

Where $\tau$ is proper time.

Since

$$d\tau^2 = g_{\alpha\beta}dx^\alpha dx^\beta \quad (\alpha,\beta = 0,1,2,3) \quad (3)$$

Then we get

$$g_{\alpha\beta} = diag(f(r), -f^{-1}(r), -r^2, -r^2\sin^2\theta) \quad (4)$$

We consider the matter configurations which do not deform (2). By considering matter energy-momentum tensor

$$T^\alpha_\beta = diag(\rho, -P_r, -P_\theta, -P_\varphi) \quad (5)$$

From equation (2) we get the components of Ricci are

$$R_{00} = -f\left(\frac{f''}{2} + \frac{f'}{r}\right)$$

$$R_{11} = f^{-1}\left(\frac{f''}{2} + \frac{f'}{r}\right)$$

$$R_{22} = f'r + f - 1$$

$$R_{33} = \sin^2\theta(f'r + f - 1) \quad (6)$$

and scalar tensor is

$$R = -f'' - \frac{4}{r}f' - \frac{2}{r^2}f + \frac{2}{r^2} \quad (7)$$

(See appendix A)

where prime (') denotes the derivative with respect to radial coordinate $r$. Substituting from equations (4) - (7) into (1) we get the equations of motion are

$$\frac{f'}{r} + \frac{f}{r^2} - \frac{1}{r^2} = -\frac{1}{l^2} + 8\pi G\rho \quad (8)$$

$$\frac{f'}{r} + \frac{f}{r^2} - \frac{1}{r^2} = -\frac{1}{l^2} - 8\pi G P_r \quad (9)$$

$$\frac{f''}{2} + \frac{f'}{r} = -\frac{1}{l^2} - 8\pi G P_\theta \quad (10)$$

$$\frac{f''}{2} + \frac{f'}{r} = -\frac{1}{l^2} - 8\pi G P_\varphi \quad (11)$$

We obtain the solutions of $f(r), \rho, P_r, P_\theta, P_\varphi$ from (8),(9),(10) and (11) we get

$$f(r) = 1 - \frac{r^2}{3l^2} + \frac{8\pi G}{r}\int_0^r \rho r^2 \, dr \quad (12)$$

$$P_r = -\rho \quad (13)$$

$$P_\theta = P_\varphi \quad (14)$$

Equations (13) and (14) show that a non-vanishing radial pressure for arbitrary matter distribution and isotropic tangential components. Note that equation (2) together with (5) determine the metric and matter's pressure in terms of $\rho$.

### III. Gaussian distribution:

Using Gaussian distribution function

(15) $\rho = \frac{A}{L^3} e^{-\frac{r^2}{L^2}}$

Where L is the characteristic length scale of the matter distribution and $A$ is the normalization constant. Substituting from (15) into (12) we get

$$(16) f(r) = 1 - \frac{r^2}{3l^2} + \frac{8\pi GA}{r} \int_0^r \frac{r^2}{L^3} e^{-\frac{r^2}{L^2}} dr$$

Performing integration, then we get

$$(17) f(r) = 1 - \frac{r^2}{3l^2} + \frac{4\pi GA}{r} \left[ \frac{\sqrt{\pi}}{2} \text{erf}\left(\frac{r}{L}\right) - \frac{r}{L} e^{-\frac{r^2}{L^2}} \right]$$

where erf(x) is the error function, defined by

$$\text{erf}(x) = \frac{2}{\sqrt{\pi}} \int_0^x e^{-t^2} dt$$

The solution of Einstein equation in the vacuum solution of Schwarzschild solution in four dimensions [9].

$$f(r) = 1 - \frac{r^2}{3l^2} - \frac{2MG}{r} \quad (18)$$

where M is the mass of black hole. Our calculation in the background of the vacuum solution ($L \to 0$), then the equation (17) becomes

$$f(r) = 1 - \frac{r^2}{3l^2} + \frac{4\pi\sqrt{\pi} GA}{r} \quad (19)$$

From equation (19) and equation (18), we get

$$A = \frac{-M}{\pi\sqrt{\pi}} \quad (20)$$

Substituting equation (20) into equation (17), we get

$$f(r) = 1 - \frac{r^2}{3l^2} - \frac{4MG}{r} \left[ \frac{1}{2} \text{erf}\left(\frac{r}{L}\right) - \frac{r}{\sqrt{\pi}L} e^{-\frac{r^2}{L^2}} \right] \quad (21)$$

We can obtain the explicit form of the mass of black hole in terms of the radial coordinate by setting $f(r) = 0$

$$M = \frac{\left(r_+ - \frac{r_+^3}{3l^2}\right)}{2G\left[\text{erf}\left(\frac{r_+}{L}\right) - \frac{2r_+}{\sqrt{\pi}L}e^{-\frac{r_+^2}{L^2}}\right]} \quad (22)$$

The extremal black hole which is the smallest black hole it is depend on the extremal radius which can be determine by $(\partial M/\partial r_H)|_{r=r_o} = 0$, where

$$2G\left(\frac{\partial M}{\partial r_H}\right)\Big|_{r=r_o} = \left\{\left(1-\frac{r_+^2}{l^2}\right)\left[\text{erf}\left(\frac{r_+}{L}\right) - \frac{2r_+}{\sqrt{\pi}L}e^{-\frac{r_+^2}{L^2}}\right] - \frac{4r_+^3}{\sqrt{\pi}L^3}\left(1-\frac{r_+^2}{3l^2}\right)e^{-\frac{r_+^2}{L^2}}\right\}\left[\text{erf}\left(\frac{r_+}{L}\right) - \frac{2r_+}{\sqrt{\pi}L}e^{-\frac{r_+^2}{L^2}}\right]^{-2} \quad (23)$$

But this is very difficult to obtain the extremal radius explicit but we can write pressure in terms of the extremal radius by

$$P = \frac{\left[\text{erf}\left(\frac{r_+}{L}\right) - \frac{2r_+}{\sqrt{\pi}L}e^{-\frac{r_+^2}{L^2}}\right] - \frac{8Gr_+^2}{\sqrt{\pi}}e^{-\frac{r_+^2}{L^2}}}{8\pi\left\{\frac{8Gr_+^2}{3\sqrt{\pi}} - \left[r_+^2\,\text{erf}\left(\frac{r_+}{L}\right) - \frac{2r_+^3}{\sqrt{\pi}L}e^{-\frac{r_+^2}{L^2}}\right]\right\}} \quad (24)$$

**IV. The thermodynamics variables:**

Now we want to obtain the equation of state we started our calculation by calculate the thermodynamics temperature (Hawking temperature) from equation (21)

$$T = \frac{\kappa}{2\pi} = \frac{f'(r)}{4\pi}\Big|_{r=r_+} \quad (25)$$

Where $\kappa$ is the surface gravity of black hole.

So,

$$T = \frac{1}{4\pi}\left\{\frac{1}{r_+} - \frac{r_+}{l^2} - \frac{4r_+}{\sqrt{\pi}L^3}\left(r_+ - \frac{r_+^3}{3l^2}\right)e^{-\frac{r_+^2}{L^2}}\left[\text{erf}\left(\frac{r_+}{L}\right) - \frac{2r_+}{\sqrt{\pi}L}e^{-\frac{r_+^2}{L^2}}\right]^{-1}\right\} \quad (26)$$

Where $\frac{d}{dx}(\text{erf}(x)) = \frac{2}{\sqrt{\pi}}e^{-x^2}$. We can easily prove that this temperature reduce to Hawking temperature of the ordinary Shwarzschild black hole in vacuum $T = \frac{1}{4\pi}\left\{\frac{1}{r_+} - \frac{r_+}{l^2}\right\}$ for $(L \to 0)$. This equation represent equation of state (see figure 3).

Now we want to calculate the thermodynamics volume as

$$V = \frac{\partial M}{\partial P}\Big|_{r=r_+} = \frac{\frac{4\pi r_+^3}{3}}{G\left[\text{erf}\left(\frac{r_+}{L}\right) - \frac{2r_+}{\sqrt{\pi}L}e^{-\frac{r_+^2}{L^2}}\right]} \quad (27)$$

From (21) we see that for $(L \to 0)$ the denominator tends to 1 which obtain the thermodynamics volume of the ordinary Schwarzschild black hole in vacuum $V = \frac{4\pi r_+^3}{3}$ (see figure 4).

The heat capacity at constant pressure is

$$C_P = \frac{\partial M}{\partial r_+}\left(\frac{\partial T}{\partial r_+}\right)^{-1}$$

$$= \frac{\left(1-\frac{r_+^2}{l^2}\right)\left[\text{erf}\left(\frac{r_+}{L}\right)-\frac{2r_+}{\sqrt{\pi}L}e^{-\frac{r_+^2}{L^2}}\right]-\frac{4r_+^3}{\sqrt{\pi}L^3}\left(1-\frac{r_+^2}{3l^2}\right)e^{-\frac{r_+^2}{L^2}}}{\frac{16r_+^4}{\pi L^5}\left(1-\frac{r_+^2}{3l^2}\right)e^{-\frac{r_+^2}{L^2}}-\frac{8r_+}{\sqrt{\pi}L^2}\left(1-\frac{2r_+^2}{3l^2}-\frac{r_+^2}{L^2}+\frac{r_+^4}{3l^2L^2}\right)e^{-\frac{r_+^2}{L^2}}\left[\text{erf}\left(\frac{r_+}{L}\right)-\frac{2r_+}{\sqrt{\pi}L}e^{-\frac{r_+^2}{L^2}}\right]-\left(\frac{1}{r_+^2}+\frac{1}{l^2}\right)\left[\text{erf}\left(\frac{r_+}{L}\right)-\frac{2r_+}{\sqrt{\pi}L}e^{-\frac{r_+^2}{L^2}}\right]^2} \quad (28)$$

**V. In general distribution form**

Now, let me consider the distribution function in general form is [8],

$$\rho = A\frac{r^n}{L^{n+3}}e^{-\frac{r^2}{L^2}} \quad (29)$$

Where L is the characteristic length scale of the matter distribution and $A$ is the normalization constant. For $n = 0$ we get Gaussian distribution, $n = 1$ Rayleigh distribution and $n = 2$ Maxwell-Boltzmann distribution, etc. substituting from (29) into (12) and integrate we get

$$f(r) = 1 - \frac{r^2}{3l^2} + \frac{4\pi GA}{r}\gamma\left(\frac{n}{2}+\frac{3}{2},\frac{r^2}{L^2}\right) \quad (30)$$

where

$$\gamma\left(\frac{n}{2}+1,x^2\right) = \int_0^{x^2} t^{\frac{n}{2}}e^{-t}dt,$$

$$\Gamma\left(\frac{n}{2}+1,x^2\right) = \int_{x^2}^{\infty} t^{\frac{n}{2}}e^{-t}dt = \Gamma\left(\frac{n}{2}+1\right) - \gamma\left(\frac{n}{2}+1,x^2\right)$$

are the incomplete lower and upper Gamma function respectively.

But in the vacuum background solution of Schwarzschild solution

$$f(r) = 1 - \frac{r^2}{3l^2} - \frac{2MG}{r} \quad (31)$$

Vacuum solution, i.e. $L \to 0$ then (30) becomes

$$f(r) = 1 - \frac{r^2}{3l^2} + \frac{4\pi GA}{r}\Gamma(\frac{n}{2}+\frac{3}{2}) \quad (32)$$

By comparing (31) and (32) then we get

$$A = -\frac{M}{2\pi\Gamma(\frac{n}{2}+\frac{3}{2})} \quad (33)$$

where it obtain the same normalization constant at n=0 eqn.(20). Then we can write (30) as

$$f(r) = 1 - \frac{r^2}{3l^2} - \frac{2MG}{r}[1 - \frac{\Gamma(\frac{n}{2}+\frac{3}{2}, \frac{r^2}{L^2})}{\Gamma(\frac{n}{2}+\frac{3}{2})}] \quad (34)$$

Now, we want to see whether there exists a black hole horizon but this located at singularity this means that $f(r) = 0$ has interior roots other than the cosmological horizons.

$$1 - \frac{r^2}{3l^2} - \frac{2MG}{r}[1 - \frac{\Gamma(\frac{n}{2}+\frac{3}{2}, \frac{r^2}{L^2})}{\Gamma(\frac{n}{2}+\frac{3}{2})}] = 0 \quad (35)$$

But it is very difficult to solve this equation analytically.

We can use the easy way to find the (interior) black hole horizon by considering the *Nariai limit* where the black hole horizon $r_-$ and the cosmological horizon $r_+$ intersect i.e. $r_- = r_+ \equiv r_{Nar}$ this happened at $f' = 0 \; and \; f = 0$. If there exist a positive solution for Nariai radius $r_{Nar}$ then there exist the black hole horizon and a cosmological horizon due to $r_- \leq r_{Nar} \leq r_+$ from (8) we get

$$r^2 = l^2[1 - \frac{4MG}{\Gamma(\frac{n}{2}+\frac{3}{2})r}\left(\frac{r}{L}\right)^{n+3} e^{-\frac{r^2}{L^2}}] \quad (36)$$

We note that for $L \to 0$ we get $r^2 = l^2$ which we get this from the ordinary Schwarzschild by using (31).

## *VI. Properties and thermodynamics of de Sitter Black hole*

I want to know the detailed form of the horizon $r_-$ to study the physical properties of de Sitter black holes since (35) which cannot be solved analytically, then we must be considered the perturbative solution near the origin by consider $r_-$ is very small this reasonable for small L

Since $r_-$ should be zero when L tends to zero by expanding (34) near $r \simeq 0$, neglecting higher-order terms.

We get equation (34) becomes

$$f(r) = 1 - \frac{r^2}{3l^2}[1 + \frac{12MGl^2}{(n+3)\Gamma(\frac{n}{2}+\frac{3}{2})}\frac{r^n}{L^{n+3}} + o(\frac{r^{n+2}}{L^{n+5}})] \quad (37)$$

(See Appendix B)

Then the black hole horizon is

$$r_- = \left[\frac{(n+3)\Gamma(\frac{n}{2}+\frac{3}{2})}{4mG}\left(1 - \frac{r^2}{3l^2}\right)\right]^{\frac{1}{n+2}} L^{1+\frac{1}{n+2}} \quad (38)$$

From this we get the black hole horizon proportional to $L$ and we get that $r_- \to 0$ as $L \to 0$ which agreed with our assumption above which provides a validity of our approximation.

Now we want to obtain the Hawking temperature

$$T_H = \frac{\hbar\kappa}{2\pi}|_{r_\pm} = \frac{\hbar L x_\pm}{6\pi l^2}|1 + \left(3\frac{l^2}{L^2} - x_\pm^2\right)\left(\frac{x_\pm^{n+1}e^{-x_\pm^2}}{\gamma(\frac{n}{2}+\frac{3}{2},x_\pm^2)} - \frac{1}{2x_\pm^2}\right)| \quad (39)$$

(See Appendix C)

where $\kappa = \left|\frac{\partial f}{2 \partial r}\right|$ is the positive surface gravity and $x_\pm = \frac{r_\pm}{L}$. From this we get this temperature is reduce to ordinary Schwarzschild temperature $T = \frac{l^2 - r_\pm^2}{4\pi l^2 r}$ at $(L \to 0)$. I driving (39) from (34) and substituting the expressed of M from (35). Though we do not know exact form of $r_\pm$ interms of M, n we note that the temperature (39) is expressed in terms of $r_\pm$. In all figures in this section, we plot the comparison for the Hawking temperature in exact form (39), approximated formula (40) and vacuum Schwarzschild solution. The right-hand side curves represent the temperature due to the cosmological horizon $r_+$. And the left-hand side curves represent those the temperature due to black hole horizon $r_-$.

Now, in order to study the thermodynamics of black holes, we must be consider the *small* black hole case, i.e. small $r_-$ which reduces the Hawking temperature as

$$T_H^- = \frac{\hbar}{4\pi r_-}\left\{(n+2) - \left[\frac{nL^2}{3l^2} + \frac{2(n+3)}{(n+5)}\right]\left(\frac{r_-}{L}\right)^2 + O\left(\left(\frac{r_-}{L}\right)^4\right)\right\} \quad (40)$$

From this we note that the temperature not vanishes for any distribution. But from this approximated temperature formula (40) we can find the approximated formula for small black hole horizon the Nariai radius $x_{Nar} = \frac{r_-}{L}$ where the temperature vanishes

$$x_{Nar} = \sqrt{\frac{3l^2(n+5)(n+2)}{n(n+5)L^2 + 6l^2(n+3)}} \quad (41)$$

Now we want to deduce the mass of the small black hole from entropy, but there is no canonical derivation of it. But we know that the entropy proportional to the area of the event horizon [10]. In other words we, assume that the black holes entropy is

$$S \equiv \alpha \frac{4\pi r_-^2}{4\hbar G} \quad (28)$$

where coefficient $\alpha$ is not fixed to one as in Bekenstien-Hawking entropy for large black hole[11] , similarly as a small black hole in higher curvature [12].

From the first law of thermodynamics

$$dM^- = T_H dS = \alpha \frac{2\pi r_-}{\hbar G} T_H dr_- \quad (42)$$

This yields mass of small black holes

$$M(r_-) = \int_0^{r_-} \alpha \frac{2\pi r_-}{\hbar G} T_H dr_-$$

$$= \alpha \frac{L^2}{3l^2 G} \int_0^{\frac{r_-}{L}} x^2 [L + (3l^2 - x^2 L^2)] \left( \frac{x^{n+1} e^{-x^2}}{\gamma(\frac{n}{2}+\frac{3}{2}, x_\pm^2) L} - \frac{1}{2x^2 L} \right) dx \quad (43)$$

We get $M^-(0) = 0$ in order to be agreed with the vacuum solution without black hole i.e. $r_- = 0$.

But we find that the analytic integration of (43) is not available. So for small black hole approximation we get, the equation (43) becomes

$$M(r_-) = \frac{\alpha r_-}{2G} \left\{ n + 2 - \frac{1}{3} \left[ \frac{nL^2}{3l^2} + \frac{2(n+3)}{(n+5)} \right] \left( \frac{r_-}{L} \right)^2 + O\left( \left( \frac{r_-}{L} \right)^4 \right) \right\} \quad (44)$$

The heat capacity $C = \frac{dM}{dT_H}$ is given by

$$C = \frac{2\alpha\pi L^2}{\hbar G} T_H x \frac{dx}{dT_H} = \frac{2\alpha\pi L^2}{\hbar G} T_H x \left(\frac{dT_H}{dx}\right)^{-1} (45)$$

$$\frac{dT_H}{dx} = \frac{1}{6L\gamma^2 l^2 x^2} |\{3(x^2 L^2 + l^2)\gamma^2 + [(6n+12)l^2 - (2n+8)L^2 x^2 - 12l^2 x^2 + 4L^2 x^4] x^{n+3} \gamma e^{-x^2} + (4L^2 x^2 - 12l^2) x^{2n+6} e^{-2x^2} \}| \quad (46)$$

But for small black hole approximation we get,

$$\frac{dT_H}{dx} = -\frac{\hbar}{4\pi L} \left\{ \frac{n+2}{x^2} + \left[ \frac{nL^2}{3l^2} + \frac{2(n+3)}{(n+5)} \right] + O((x)^4) \right\} (47)$$

## *VII. Conclusions*

In this paper, we obtain the solution of Einstein equation in smearing of point matter by consider Gaussian distribution function, then by using the mass of black hole in terms of outer radius as enthalpy we can obtain some thermodynamics quantity like (Hawking temperature, thermodynamics volume and heat capacity). We note that every quantity reduce to the quantity of ordinary Schwarzschild black hole (vacuum solution) as $L \to 0$ . In figure 1 we plot $f(r)$ with the radial coordinate and we note that its goes to zero for large pressure at infinity of $r$ and for $r \to 0$ it goes to mins infinity. In figure2 we plot the mass of black hole with r and we note that for very small radial the mass tends to zero. In figure 3 we plot the Hawking temperature with $r$ and we note that the temperature must be positive as the radial less than the extremal radius. In figure 4 we plot the thermodynamics volume with $r$ and we note that the volume become zero for small radius and it is not depends on pressure. In second we considered the general form of distribution function, including Gaussian distribution at $n = 0$ Rayleigh distribution at $n = 1$ and Maxwell-Boltzmann distribution at $n = 2$. We study the physical properties of black holes by obtaining the explicit form of the black holes horizon by expanding equation (36) near $r \approx 0$ . We are studying the thermodynamics properties of a black hole by using the relation between the entropy and the area and then obtain the mass of the small black hole from the first law of thermodynamics. Finally, we plot some figures between Hawking temperature and Nariai radius in figures. 5, 6, 7 and 8 for $n = 0, 1, 2 \ and \ 3$ respectively, which represent comparison between the exact temperature (39), approximation temperature (40) and the vacuum solution in four dimension but in figures 9 and 10 we plot the exact temperature and approximate temperature respectively which corresponding to Gaussian distribution at $n = 0$ Rayleigh distribution at $n = 1$ and Maxwell-Boltzman distribution at $n = 2$. We note that the two temperatures are identical for the black hole horizon which is less than the Nariai radius.

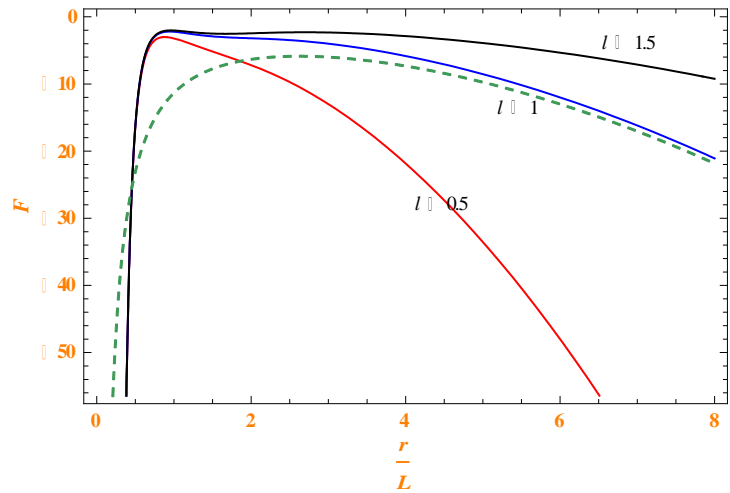

Figure 1: The relation between $f(r)$ and $\frac{r}{L}$ by varying pressure where the red solid line corresponding $l = 0.5$, blue line corresponding to $l = 1$, the black line corresponding to $l = 1.5$ and dashed line represent the mass of ordinary Shwarzschild black hole.

.

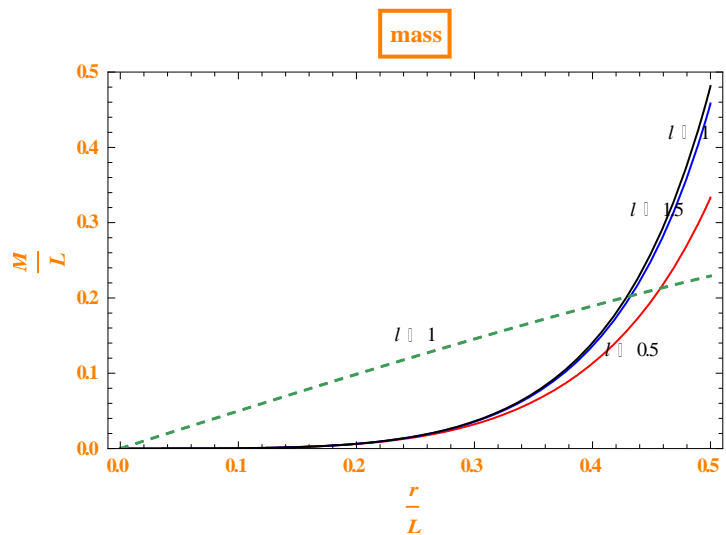

Figure 2: The relation between $M$ and $\frac{r}{L}$ by varying pressure where the red solid line corresponding to $l = 0.5$ , blue line corresponding to $l = 1$ and the black line corresponding to $l = 1.5$ and dashed line represent the mass of ordinary Shwarzschild black hole .

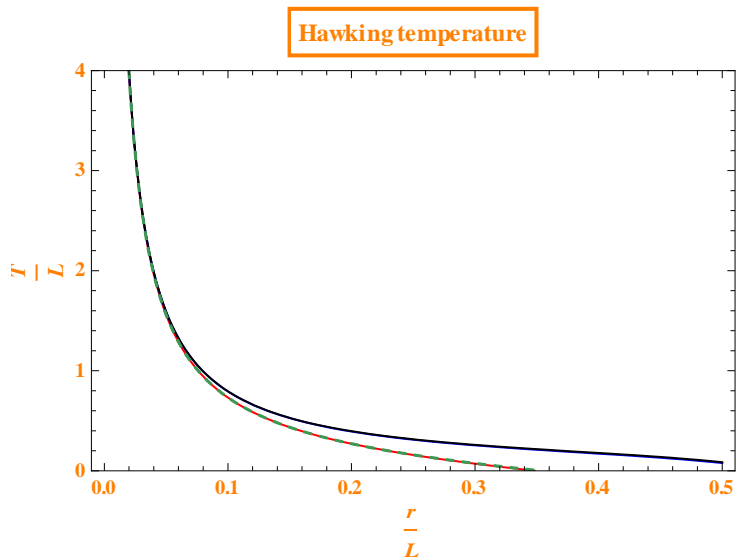

Figure 3. The Hawking temperature with radial coordinate the red solid line refer to $l = 0.5$ , blue line corresponding to $l = 2.5$ and dashed line represent the temperature of ordinary Shwarzschild black hole

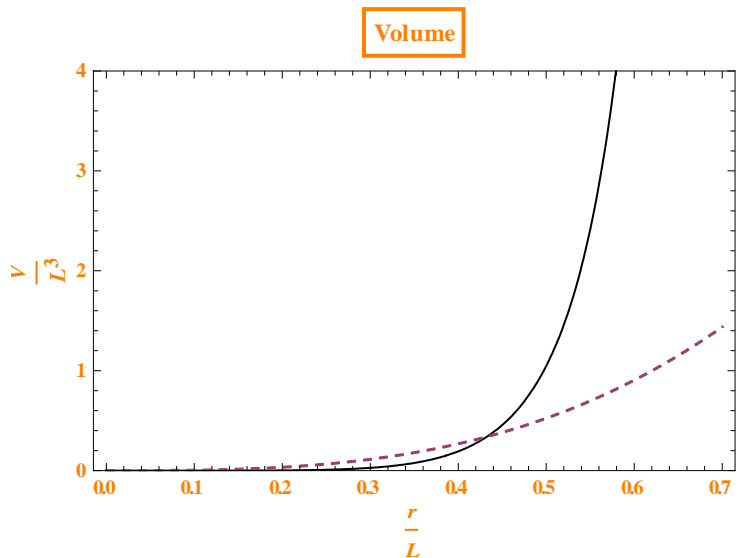

Figure 4. The volume with the radial coordinate where the solid line represents (27) and dashed line corresponding Schwarzschild volume.

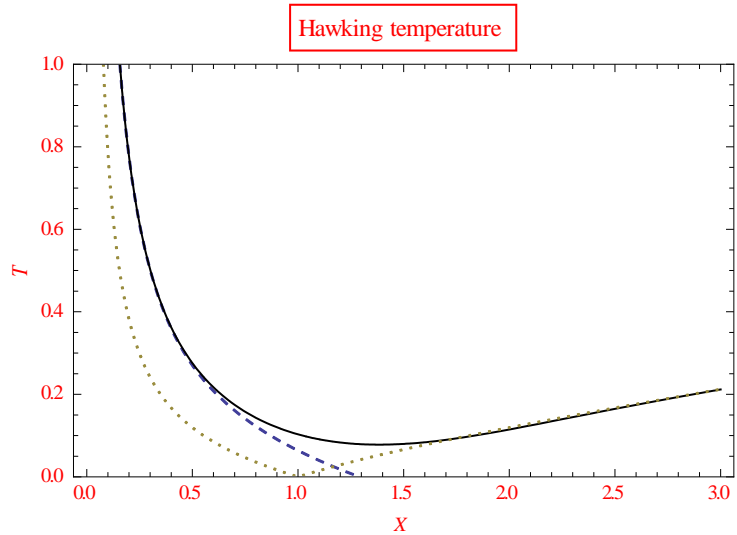

Figure 5 The Hawking temperature vs. $x = \frac{r_\pm}{L}$ for $n = 0$ (Gaussian distribution) from left to right where the solid line donated exact temperature (39), the dashed line donated the approximated temperature (40) and dotted line donated the temperature for vacuum in four dimension where putting $\hbar = L = l \equiv 1$

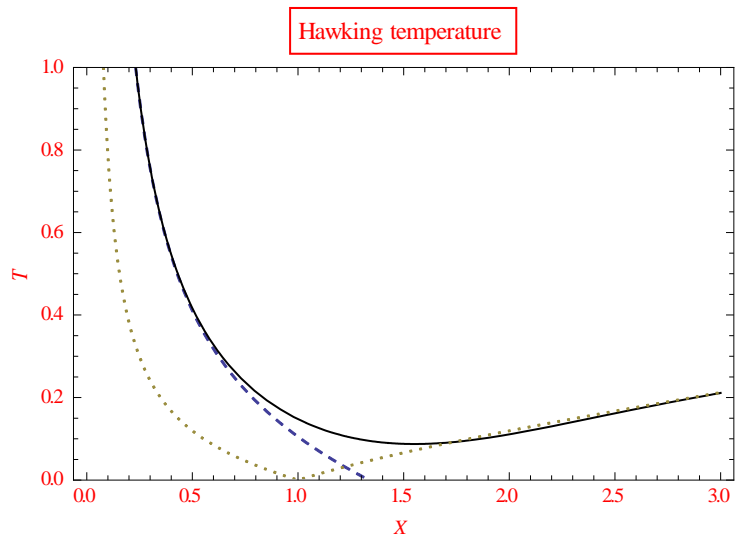

Figure 6 The Hawking temperature vs. $x = \frac{r_\pm}{L}$ for $n = 1$ (Rayleigh distribution) from lift to right where the solid line donated exact temperature (39), the dashed line donated the approximated

temperature (40) and dotted line donated the temperature for vacuum in four dimension where putting $\hbar = L = l \equiv 1$

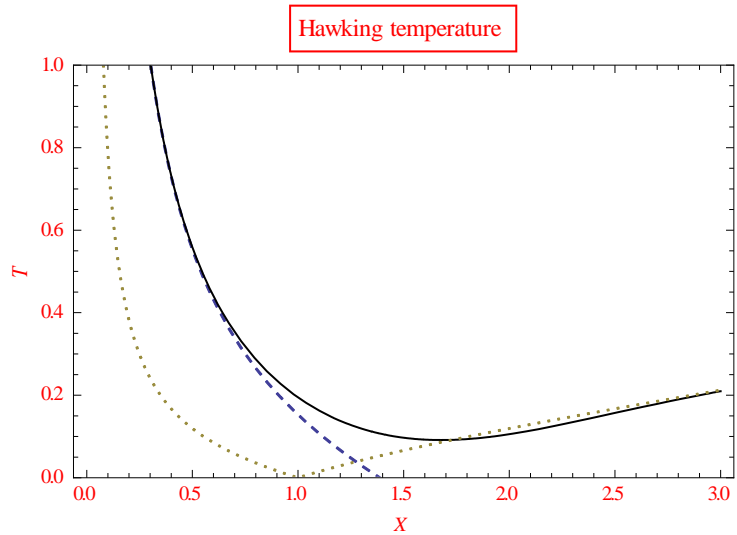

Figure 7 The Hawking temperature vs. $x = \frac{r_\pm}{L}$ for $n = 2$ (Maxwell-Boltzmann) distribution from lift to right where the solid line donated exact temperature (39), the dashed line donated the approximated temperature (40) and dotted line donated the temperature for vacuum in four dimension where putting $\hbar = L = l \equiv 1$

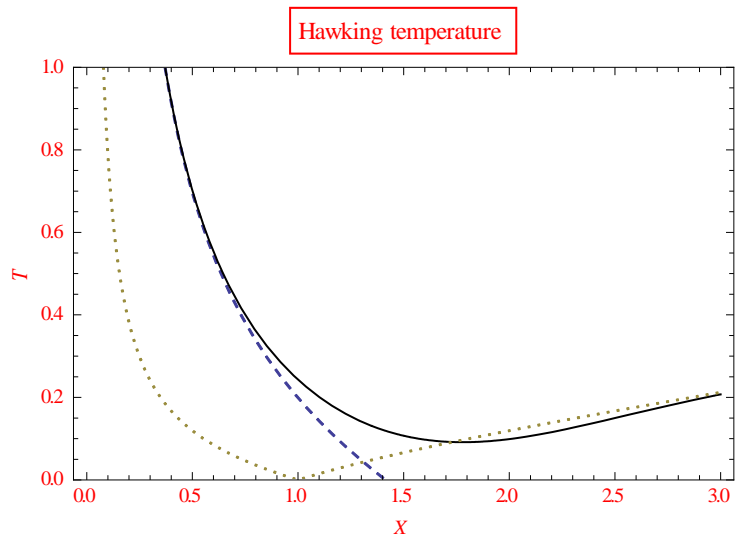

Figure 8 The Hawking temperature vs. $x = \frac{r_\pm}{L}$ for $n = 3$ from lift to right where the solid linedonated exact temperature (39), the dashed line donated the approximated temperature (40) and dotted line donated the temperature for vacuum in four dimension where putting $\hbar = L = l \equiv 1$

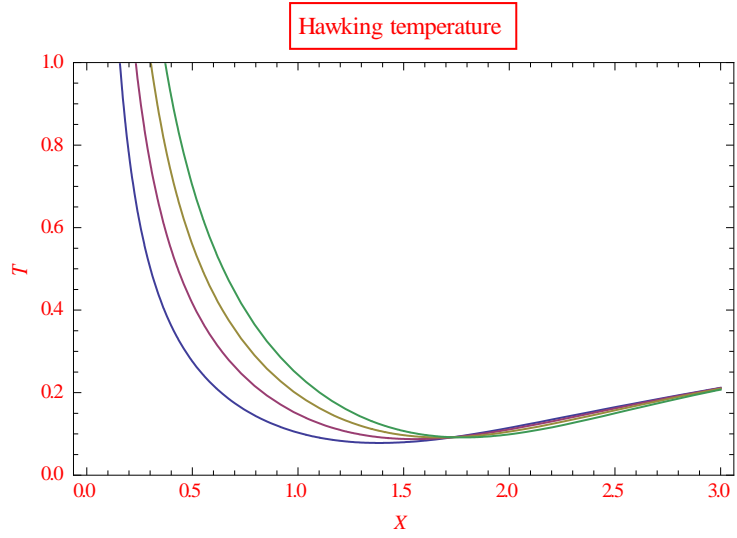

Figure 9 The Hawking temperature vs. $x = \frac{r_\pm}{L}$ for $n = 0,1,2,3$ from bottom curves to top curves respectively. For exact temperature (39), where putting $\hbar = L = l \equiv 1$

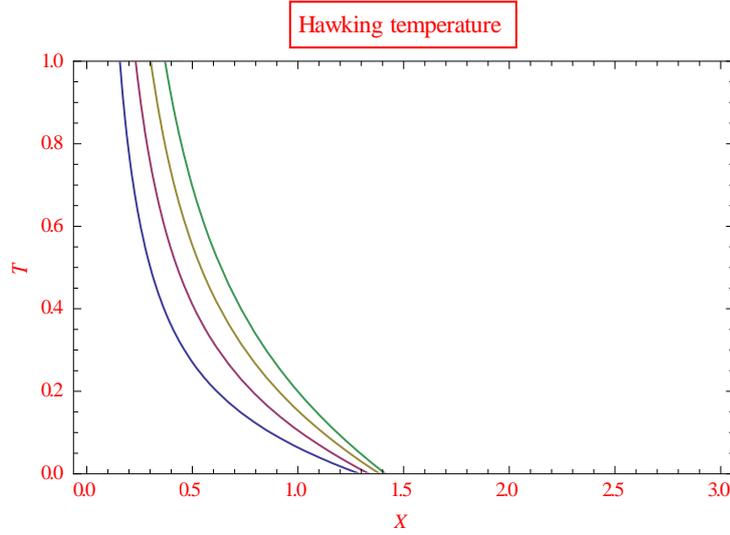

Figure 10 The Hawking temperature vs. $x = \frac{r_\pm}{L}$ for $n = 0,1,2,3$ from lift curves to right curves respectively. For approximate temperature (40), where putting $\hbar = L = l \equiv 1$

**Appendix A**

From equation (2), we get the Lagrangian is

$$L(\dot{x}^\sigma, x^\sigma) = f(r)\dot{t}^2 - \frac{1}{f(r)}\dot{r}^2 - r^2\dot{\theta}^2 - r^2\sin^2\theta\,\dot{\varphi}^2 \quad \text{(A1)}$$

where dots denote differentiation with respect to an affine parameter $u$. Partial differentiation gives:

$$\frac{\partial L}{\partial \dot{t}} = f(r)\dot{t}$$

$$\frac{\partial L}{\partial \dot{r}} = -f^{-1}\dot{r}$$

$$\frac{\partial L}{\partial \dot{\theta}} = -r^2\dot{\theta}$$

$$\frac{\partial L}{\partial \dot{\varphi}} = -2r^2\sin^2\theta\,\dot{\varphi}$$

$$\frac{\partial L}{\partial t} = 0 \quad \text{(A2)}$$

$$\frac{\partial L}{\partial t} = 0$$

$$\frac{\partial L}{\partial r} = \frac{1}{2}f'(r)\dot{t}^2 + \frac{f'(r)}{2f^2(r)}\dot{r}^2 - r^2\dot{\theta}^2 - r^2\sin^2\theta\,\dot{\varphi}^2$$

$$\frac{\partial L}{\partial \theta} = -r^2\sin\theta\cos\theta\,\dot{\varphi}^2$$

$$\frac{\partial L}{\partial t} = 0$$

Substitution from (A2) into Euler-Lagrange equations $\frac{d}{du}\left(\frac{\partial L}{\partial \dot{x}^a}\right) - \frac{\partial L}{\partial x^a} = 0 \ (a = 0,1,2,3)$, we get the equations of motion are

$$\ddot{t} + f'f^{-1}\dot{r}\dot{t} = 0$$

$$\ddot{r} + \frac{1}{2}f'f\,\dot{t}^2 - \frac{1}{2}f'f^{-1}\dot{r}^2 - rf\,\dot{\theta}^2 - rf\,\sin^2\theta\,\dot{\varphi}^2 = 0$$

$$\ddot{\theta} + \frac{2}{r}\dot{r}\dot{\theta} - \sin\theta\cos\varphi\,\dot{\varphi}^2 = 0$$

$$\ddot{\varphi} + \frac{2}{r}\dot{r}\dot{\varphi} + 2\cot\theta\,\dot{\theta}\dot{\varphi} = 0 \quad (A3)$$

Comparing this equations with geodesic equations ($\ddot{x}^c + \Gamma^c_{ab}\dot{x}^a\dot{x}^b = 0$) we get the nine independent non-zero connection coefficients in the following form

$$\Gamma^0_{01} = \frac{1}{2}f'f^{-1}$$

$$\Gamma^1_{00} = \frac{1}{2}f'f$$

$$\Gamma^1_{11} = -\frac{1}{2}f'f^{-1}$$

$$\Gamma^1_{33} = -rf\,\sin^2\theta$$

$$\Gamma^2_{12} = \frac{1}{r}$$

$$\Gamma^2_{33} = -\sin\theta\cos\theta$$

$$\Gamma^3_{13} = \frac{1}{r}$$

$$\Gamma^3_{23} = \cot\theta \quad (A4)$$

The non-zero connection coefficients can be used to determine the non-zero components of the Riemann curvature tensor,
Using the general formula

$$R^{\rho}{}_{\sigma\mu\nu} = \frac{\partial \Gamma^{\rho}{}_{\sigma\nu}}{\partial x^{\mu}} - \frac{\partial \Gamma^{\rho}{}_{\sigma\mu}}{\partial x^{\nu}} + \Gamma^{\alpha}{}_{\sigma\nu}\Gamma^{\rho}{}_{\alpha\mu} - \Gamma^{\alpha}{}_{\sigma\mu}\Gamma^{\rho}{}_{\alpha\nu} \quad (A5)$$

We can obtain the Ricci tensor by contracting the Riemann tensor, gives the components

$$R_{\mu\nu} = R^{\rho}{}_{\sigma\mu\rho} \quad (A6)$$

then we get the only four diagonal components of the Ricci tensor are not identically zero:

$$R_{00} = -f\left(\frac{f''}{2} + \frac{f'}{r}\right)$$

$$R_{11} = f^{-1}\left(\frac{f''}{2} + \frac{f'}{r}\right)$$

$$R_{22} = f'r + f - 1$$

$$R_{33} = \sin^2\theta(f'r + f - 1) \quad (A7)$$

and scalar tensor is

$$R = g^{\mu\nu}R_{\mu\nu} = g^{00}R_{00} + g^{11}R_{11} + g^{22}R_{22} + g^{33}R_{33} \quad (A8)$$

So,

$$R = -f'' - \frac{4}{r}f' - \frac{2}{r^2}f + \frac{2}{r^2} \quad (A9)$$

**Appendix B**

Since

$$\Gamma\left(\frac{n}{2} + \frac{3}{2}, \frac{r^2}{L^2}\right) = \Gamma\left(\frac{n}{2} + \frac{3}{2}\right) - \gamma\left(\frac{n}{2} + \frac{3}{2}, \frac{r^2}{L^2}\right) \quad (B1)$$

So,

$$1 - \frac{\Gamma\left(\frac{n}{2} + \frac{3}{2}, \frac{r^2}{L^2}\right)}{\Gamma\left(\frac{n}{2} + \frac{3}{2}\right)} = 1 - 1 + \frac{\gamma\left(\frac{n}{2} + \frac{3}{2}, \frac{r^2}{L^2}\right)}{\Gamma\left(\frac{n}{2} + \frac{3}{2}\right)} = \frac{\gamma\left(\frac{n}{2} + \frac{3}{2}, \frac{r^2}{L^2}\right)}{\Gamma\left(\frac{n}{2} + \frac{3}{2}\right)} \quad (B2)$$

And by expanding $\gamma\left(\frac{n}{2} + \frac{3}{2}, \frac{r^2}{L^2}\right)$ near $r = 0$ then we get

$$\gamma\left(\frac{n}{2}+\frac{3}{2},\frac{r^2}{L^2}\right) = \int_0^{\frac{r^2}{L^2}} t^{\frac{n+1}{2}} e^{-t} dt = \int_0^{\frac{r^2}{L^2}} t^{\frac{n+1}{2}}\left(1 - t + \frac{t^2}{2} - \frac{t^3}{6} + \cdots\right) dt$$

$$= \int_0^{\frac{r^2}{L^2}} \left[t^{\frac{n+1}{2}} - t^{\frac{n+3}{2}} + \frac{t^{\frac{n+5}{2}}}{2} + \cdots\right] dt$$

$$= \left[\frac{2}{n+3}\left(\frac{r^2}{L^2}\right)^{\frac{n+3}{2}} - \frac{2}{n+5}\left(\frac{r^2}{L^2}\right)^{\frac{n+5}{2}} + \frac{2}{n+7}\frac{\left(\frac{r^2}{L^2}\right)^{\frac{n+7}{2}}}{2} + \cdots\right]$$

$$= \frac{2}{n+3}\left(\frac{r}{L}\right)^{n+3} + o\left(\frac{r^{n+5}}{L^{n+5}}\right) \quad (B3)$$

Substituting from (B3) into (B2), we get

$$1 - \frac{\Gamma\left(\frac{n}{2}+\frac{3}{2},\frac{r^2}{L^2}\right)}{\Gamma\left(\frac{n}{2}+\frac{3}{2}\right)} = \frac{1}{\Gamma\left(\frac{n}{2}+\frac{3}{2}\right)}\left[\frac{2}{n+3}\left(\frac{r}{L}\right)^{n+3} + O\left(\frac{r^{n+5}}{L^{n+5}}\right)\right] \quad (B4)$$

Then substituting from (B4) into (37) we get

$$f(r) = 1 - \frac{r^2}{3l^2}\left[1 + \frac{12mGl^2}{(n+3)\Gamma\left(\frac{n}{2}+\frac{3}{2}\right)}\frac{r^n}{L^{n+3}} + O\left(\frac{r^{n+3}}{L^{n+5}}\right)\right]$$

## Appendix C

Since the surface gravity $\kappa = \left|\frac{\partial f}{2\partial r}\right|$ from equation (34) we get

$$\kappa = \left|\frac{-r}{3l^2} - \frac{mG}{\Gamma\left(\frac{n}{2}+\frac{3}{2}\right)}\left[\frac{\gamma'}{r} - \frac{\gamma}{r^2}\right]\right| \quad (C1)$$

and from definition of gamma

$$\gamma' = \frac{2}{L}\left(\frac{r}{L}\right)^{n+2} e^{-\frac{r^2}{L^2}} \quad (C2)$$

Substituting from (C2) into (C1) we get

$$\kappa = \left|\frac{r}{3l^2} + \frac{mG}{\Gamma\left(\frac{n}{2}+\frac{3}{2}\right)}\left[\frac{\frac{2}{L}\left(\frac{r}{L}\right)^{n+2} e^{-\frac{r^2}{L^2}}}{r} - \frac{\gamma}{r^2}\right]\right| \quad (C3)$$

Substituting from (C3) into (39) and putting $x_\pm = \frac{r_\pm}{L}$ we get

$$T_H = \frac{\hbar \kappa}{2\pi}\bigg|_{r_\pm} = \frac{\hbar L x_\pm}{6\pi l^2}\left|1 + \left(3\frac{l^2}{L^2} - x_\pm^2\right)\left(\frac{x_\pm^{n+1} e^{-x_\pm^2}}{\gamma\left(\frac{n}{2}+\frac{3}{2}, x_\pm^2\right)} - \frac{1}{2x_\pm^2}\right)\right|$$